\documentclass[10pt]{iopart}
\usepackage{graphicx}
\usepackage{epsfig}
\usepackage{dcolumn}
\usepackage{bm}
\usepackage{amssymb}
\usepackage{ulem}

\begin{document}
	
\title[Friction Fluctuations of Gold Nanoparticles in the Superlubric Regime]{Friction Fluctuations of Gold Nanoparticles in the Superlubric Regime}

\author{Dirk Dietzel$^{1}$, Astrid S. de Wijn$^{2,3}$, Matthias Vorholzer$^{1}$, Andre Schirmeisen$^{1}$} 

\address{$^1$Institute of Applied Physics, Justus-Liebig-Universit\"at Giessen, 35392 Giessen, Germany} 
\address{$^2$Department of Mechanical and Industrial Engineering, Norwegian University of Science and Technology, 7491 Trondheim, Norway}
\address{$^3$Department of Physics, Stockholm University, 10691 Stockholm, Sweden}

\ead{schirmeisen@uni-giessen.de}
\ead{astrid.dewijn@ntnu.no}

\begin{abstract}
	
Superlubricity, or alternatively termed structural (super)lubrictiy, is a concept where ultra-low friction is expected at the interface between sliding surfaces if these surfaces are incommensurate and thus unable to interlock. In this work, we now report on sudden, reversible, friction changes that have been observed during AFM based nanomanipulation experiments of gold nanoparticles sliding on highly oriented pyrolythic graphite. These effects are can be explained by rotations of the gold nanoparticles within the concept of structural superlubricity, where the occurrence of ultra-low friction can depend extremely sensitively on the relative orientation between the slider and the substrate. From our theoretical simulations it will become apparent how even miniscule magnitudes of rotation are compatible to the observed effects and how size and shape of the particles can influence the dependence between friction and relative orientation. 
	
\end{abstract}


\submitto{\NT}
\ioptwocol

\maketitle

\section{Introduction}

Achieving ultralow friction without the need of conventional lubricants can be considered as one of the most alluring prospects in current tribology. In recent years, tribologists have come closer to this goal, as a new route was outlined by the concept of 'superlubricity' \cite{Hirano1990,Hirano1991}. Here, ultralow friction is anticipated in case of a structural mismatch at the interface between substrate and slider. This basic idea of 'superlubricity' becomes most apparent by looking at the interaction between two identical surfaces: If these are oriented in the same way, an effective interlocking of the atomic potentials can occur, which results in a distinct corrugation of the potential energy during relative motion. Without such interlocking, the corrugation will be much smaller and friction will be very low, i.e. superlubric. A configuration under which the interfaces can interlock is commonly denoted as 'commensurate', whereas interfaces without interlocking effects are typically described as 'incommensurate' and transitions between these states can be achieved by simply rotating the slider relative to the substrate.

Based on this concept a number of experimental works has so far been published where structural lubricity has been observed for material combinations like Mica on Mica \cite{Hirano1991,Hirano1993} or graphite on graphite \cite{Dienwiebel2004}. For the latter combination, often experimental configurations have been chosen, where superlubricity could be analyzed by shearing lithographically prepared stacks of graphite \cite{Feng2013,Liu2012,Vu2016,Koren2015}. Only recently these concepts of superlubricity could be applied to a macroscale systems, where ultralow friction was observed in tribometer experiments for graphene covered surfaces \cite{Berman2015}.

However, in many technical applications interfaces are formed by two different constituents. But so far, considerably less research on structural lubricity was done for such systems. This is mostly due to the fact that experimentalists often face fundamental problems in preparing and analyzing suitable interfaces, which must be atomically flat and clean. Most prominently, friction force microscopy (FFM)\cite{Mate1987}, as a primary tool in nanotribology, is usually inapt for the analysis of structural lubricity because of the ill defined interfaces between AFM tips and substrates \cite{Schirmeisen2009,Dietzel2014}, a problem that is usually only overcome by using specially terminated AFM-tips \cite{Dienwiebel2004,Liu2017}. As an alternative, FFM based nanomanipulation techniques  \cite{Luethi1994,Dietzel2008} have proven to be useful for the analysis of structural lubricity, since atomically flat and clean interfaces are often found for nanoparticle systems upon careful preparation. The interfacial friction between particles and substrate then becomes accessible by measuring the additional torsion of an AFM cantilever when pushing a nanoparticle \cite{Dietzel2014}. In recent years, such nanomanipulation based approaches have been established as a valuable addition to conventional techniques in tribology and nanotribology with special focus on analyzing superlubricity \cite{Schirmeisen2009, Dietzel2014}.

One key advantage of these nanomanipulation techniques is the fact, that nanoparticles of different size can easily be prepared. This allows to analyze the contact area dependence of friction, which can be considered a unique fingerprint of strucutral lubricity: Theory predicts that only for commensurate interfaces a linear increase of friction with contact area can be expected, while any incommensurate configuration should show a sublinear increase of the friction $F$ with area \cite{Mueser2001,deWijn2012,Mueser2007}. In this case, friction vs. area scaling can typically be described by a power law $F = F_0 \cdot N^\gamma$.  If one or both surfaces are amorphous, theory predicts a uniform scaling factor $\gamma=0.5$~(see Ref.~\cite{Mueser2001}). But for the case of crystalline interfaces, the interface conditions become more complex. Here, the actual value of $\gamma$ is in the range of $0\leq\gamma\leq 0.5$ and depends sensitively on shape, structure, relative orientation of the interface. By now, a growing number of experimental studies has confirmed this sublinear friction vs. area scaling \cite{Dietzel2013, Koren2015, Kawai2016, Cihan2016, Dietzel2017, Ozogul2017}. But so far, the influence of relative alignment on friction (and thus on $\gamma$) has rarely been considered for experiments on heterogenous interfaces.    

Here, we report on nanomanipulation experiments performed for Gold nanoparticles sliding on  highly oriented pyrolythic graphite (HOPG), a system for which a distinct sublinear friction vs. area scaling was previously found \cite{Dietzel2013,Cihan2016}. Our new analysis has now revealed the occurrence of additional friction features in the lateral force signals during particle manipulation. More specifically, we found that the ultra-low friction level of structural lubricity is often overlaid by sudden and reversible friction changes.

We will demonstrate that these fluctuations can be related to recent theoretical studies~\cite{deWijn2012, Hod2013, Leven2013}.
These show how the relative orientation between surfaces can still affect friction even if the surfaces do not share the same lattice constants or even lattice structure.
To better assess how these concepts can be reconciled with the specific friction fluctuations observed in our experiments, we have performed numerical simulations for gold nanoparticles on HOPG, that give new insight into the intricate relation between friction, particle shape, contact area, and orientation. Based on these simulations, we will show why already miniscule rotations, suitable to match our experiments, are sufficient to explain the observed friction changes and how the complex pattern of distinct orientations can prevent substantial particle rotation.  

\section{Sample System and Experimental Approach}

\begin{figure}

\includegraphics[width=0.6\linewidth,keepaspectratio]{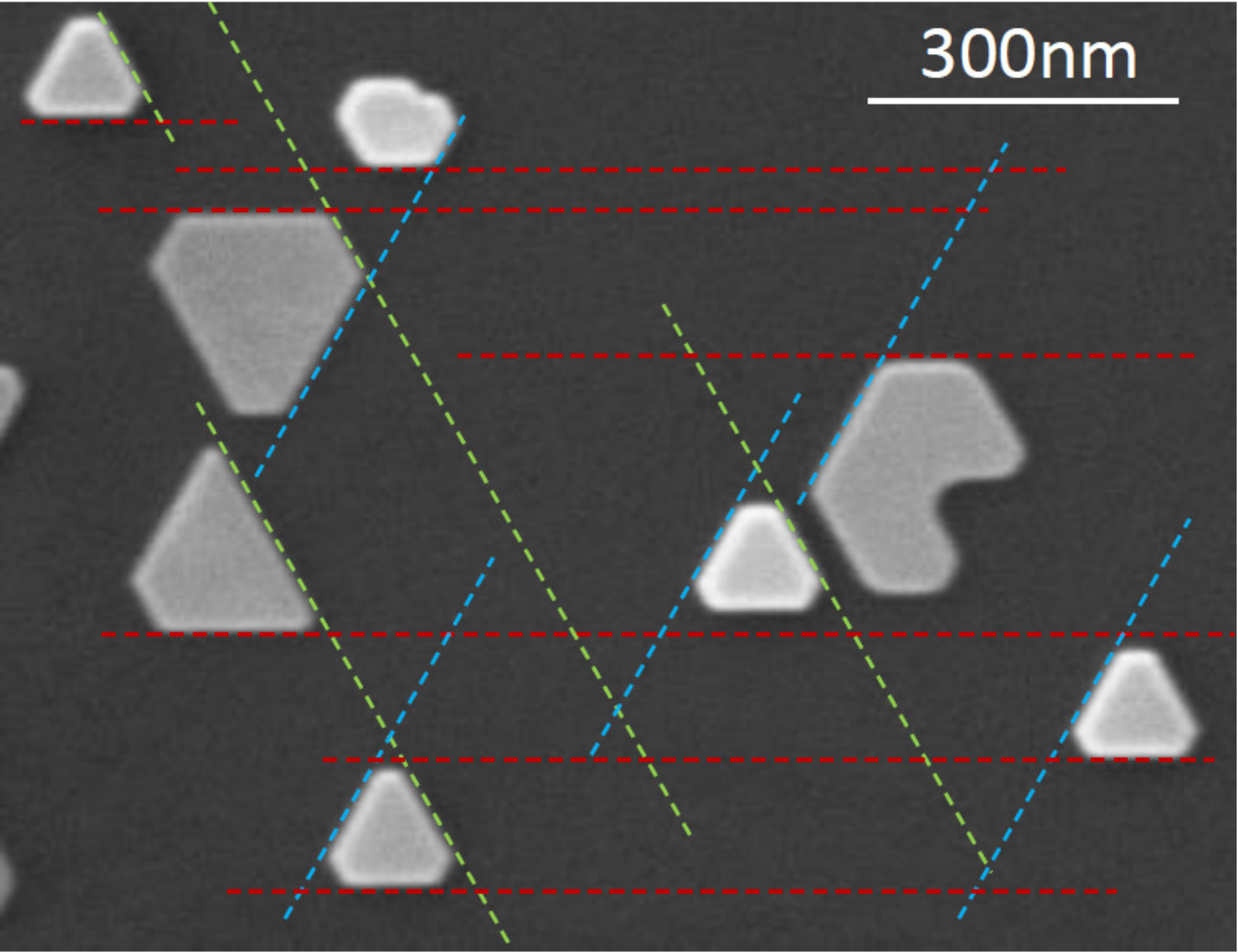}
\caption{Scanning electron microscopy image of gold Nanoparticles evaporated onto an HOPG substrate. The cristalline structure of the nanoparticles is evident from their well defined geometrical shape. Almost all nanoparticles are aligned in the same way, which indicates how the orientation of the substrate influences the growth of the nanoparticles (To emphasize the accuracy of alignment between the nanoparticles, dashed guidelines have been added to the figure).}
\label{fig:SEMAuonHOPG}
\end{figure}

Gold nanoparticles on HOPG have been chosen as a model system to analyze interfacial friction between well defined interfaces. The gold particles have been prepared by thermal evaporation onto freshly cleaved HOPG. The substrate was kept at room temperature, while the evaporation was carried out $1350^{\circ}$C. Fig.~\ref{fig:SEMAuonHOPG} shows a scanning electron microscopy (SEM) image of an ensemble of particles obtained after a typical evaporation time of 2 min, when optimal coverage of the surface was achieved.

The crystallinity of the nanoparticles is directly evident from their well-defined geometrical shape. Furthermore, it is noticeable that all nanoparticles are aligned in a similar way, which indicates how the orientation of the substrate influences the growth of the nanoparticles. The sixfold symmetry found for the preferred particle orientations (cf guide lines in Fig.~\ref{fig:SEMAuonHOPG}) is consistent with the sixfold symmetry of the nanocrystals and in good agreement with the angular difference of $60^{\circ}$ found previously between the main commensurate orientations for crystalline gold on graphite\,\cite{deWijn2012}. Intermediate commensurate orientations, which are also theoretically anticipated, have not been found for the gold nanoparticles prepared by evaporation.

From Fig.~\ref{fig:SEMAuonHOPG} it becomes immediately clear that the system cannot be viewed as isotropic, but instead the existence of distinct preferred directions has to be taken in account. Moreover, since the growth process of the nanoparticles will favor a configuration representing a potential minimum, it can be assumed that particles can initially be found in commensurate orientations and nanomanipulation experiments should in principle be suitable to record the switch from commensurate to incommensurate orientations.

\begin{figure}
	\centering
	\includegraphics[width=0.8\linewidth,keepaspectratio]{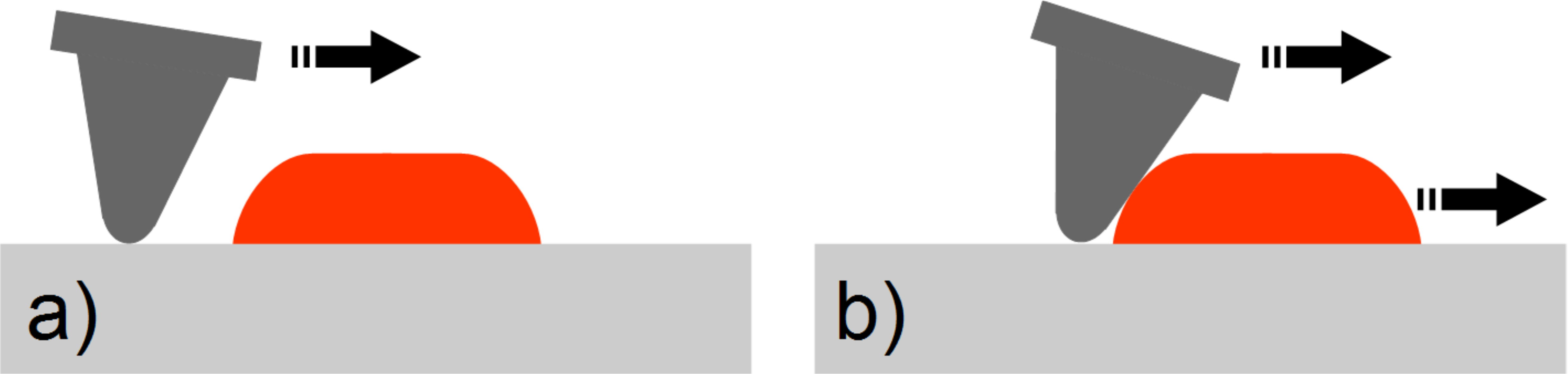}
	\caption{Concept of nanoparticle manipulation. a) While the tip of an AFM approaches the nanoparticle from the side, the torsion of the cantilever, is proportional to the friction between the particle and the substrate. b) Once the tip reaches the particle and the pushing process starts, an increased torsion of the cantilever can be measured, which is then a measure for the combined tip/substrate and particle/substrate friction contributions.}
	\label{fig:Concept}
\end{figure}

In order to assess the interfacial friction between nanoparticles and substrate, we employed contact mode nanomanipulation procedures\,\cite{Dietzel2007,Dietzel2010,Dietzel2013}. The manipulation process consists of several consecutive steps: First, a nanoparticle suitable for manipulation must be chosen. Therefore the sample is imaged using a dynamic AFM Mode (typically CE-Mode \cite{Ueyama1998,Schirmeisen2005}), which prevents unwanted nanoparticle manipulation during scanning \cite{Dietzel2010}. After a nanoparticle suitable for manipulation is found, the tip is positioned aside the nanoparticle and the AFM is switched to contact mode. Then the tip is moved along a straight path perpendicular to the cantilever axes meanwhile pushing the nanoparticle (see Fig.~\ref{fig:Concept}). If the tip path is directed through the center of mass of the nanoparticle, a straight manipulation of the particle can usually be achieved, while at the same time the nanoparticle still has some freedom of rotation. The lateral force signal is recorded during the whole tip path and the difference between the lateral force signal before and during the pushing process constitutes the interfacial friction between substrate and particle\,\cite{Dietzel2010,Dietzel2008, Dietzel2007}. After a manipulation process is finished, the AFM is switched back to CE-mode and a control image of the nanoparticle is taken. All measurements were performed using a standard Omicron STM/AFM combination under UHV conditions at room temperature. Standard Si cantilevers with force constants of appr. 0.05 N/m have been used ($\mu$masch CSC 37). A comprehensive description of the manipulation approach can be found in previous publications \cite{Dietzel2014, Dietzel2010}.

\begin{figure}
	\centering
	\includegraphics[width=0.9\linewidth,keepaspectratio]{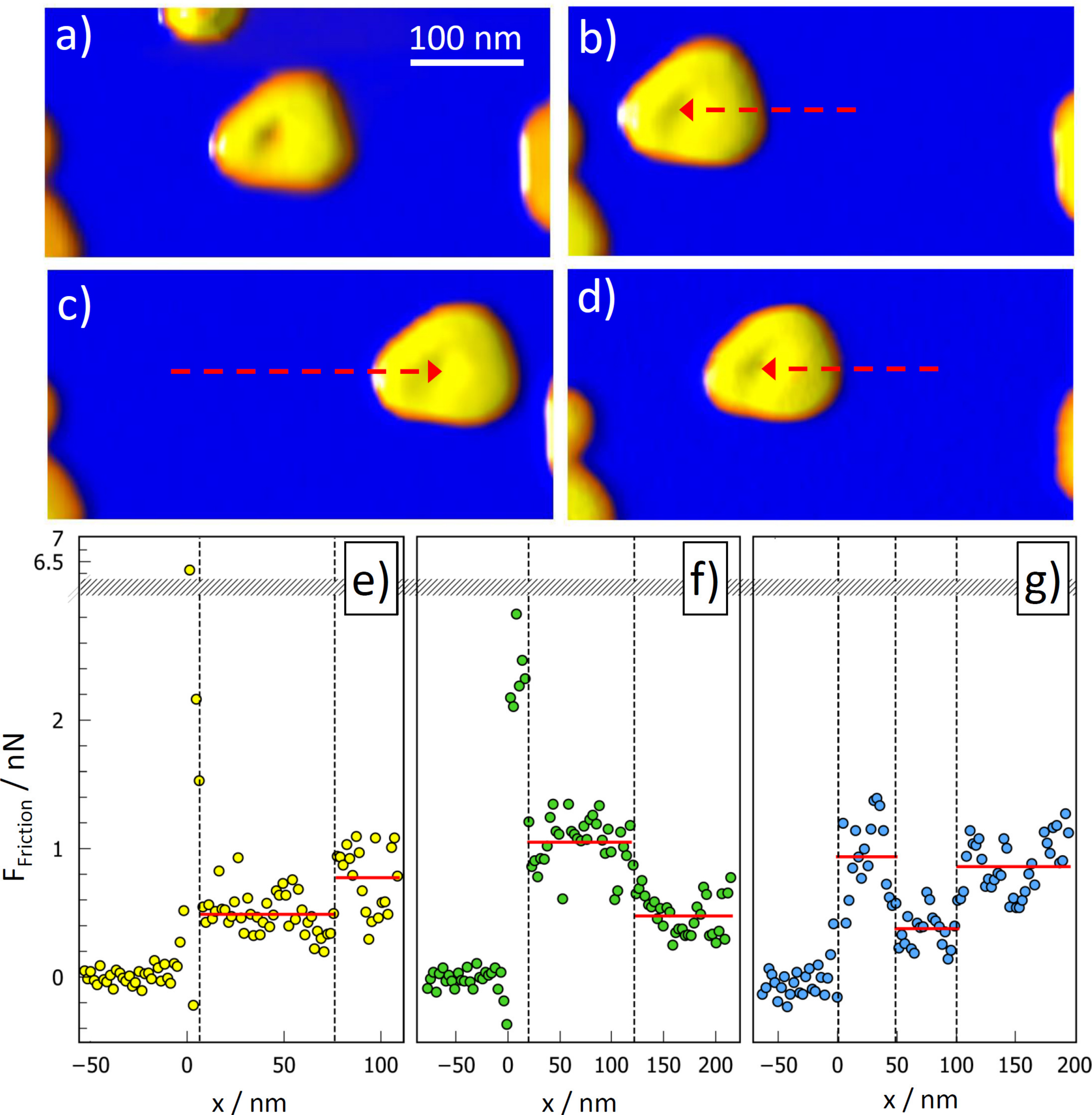}
	\caption{Normalized topography images and normalized friction traces for multiple manipulations of a single gold particle with a contact area of $A_\mathrm{contact}=12600$~nm$^2$. a) Initial position of the particle after evaporation. b)-d) resting positions after the 1st, 2nd, and 3rd manipulation. The arrows in a)-c) indicate the particle path during manipulation and the corresponding lateral force signals are shown in e)-g). In each case, the lateral force signal recorded before the tip hits the particle has been used as reference level that has been subtracted from the friction trace. Please note, that an pronounced friction peak was observed in e) and f) when sliding was initiated. To accommodate this peak in the images, we have added a break on all y-scales of e)-g).
	}
	\label{fig:Correlation}
\end{figure}

\section{Experimental Results}

While the general contact area dependence of friction for gold nanoparticles sliding on HOPG was already analyzed in a previous publication \cite{Dietzel2013}, we have now focussed on the friction fluctuations, which can be observed within the lateral force signals recorded during particle movement.   

As an example, Fig.~\ref{fig:Correlation}a shows a gold nanoparticle that appears to be of triangular shape with rounded corners at its initial position. The particle has been pushed by the cantilever three times and the respective resting positions are shown in Fig.~\ref{fig:Correlation}b-d. Interestingly, the corresponding lateral force signals reveal significant but reversible changes in friction during the manipulation process, as it is shown in Fig.~\ref{fig:Correlation}~e-g. 

For each of theses manipulations, we used the average friction level before the tip reaches the nanoparticle as reference, resulting in a lateral force signal around zero before tip/particle contact ($x<0$ in Fig.~\ref{fig:Correlation}e-g). Any subsequent increase in normalized friction can thus be directly attributed to effects at the interface between nanoparticle and substrate. During the different manipulations, the friction level changes several times, with periods of relative stability limited to sliding distances in the range of $50-100$nm, as indicated in Fig. \ref{fig:Correlation}e-g. The friction changes thus appear to be at seemingly random position, but generally alternating between higher and lower friction values. Additionally, two pronounced peaks in the lateral force signal can be observed, when particle sliding is initiated in Fig. \ref{fig:Correlation}e and f.

\begin{figure}
	\includegraphics[width=1\linewidth,keepaspectratio]{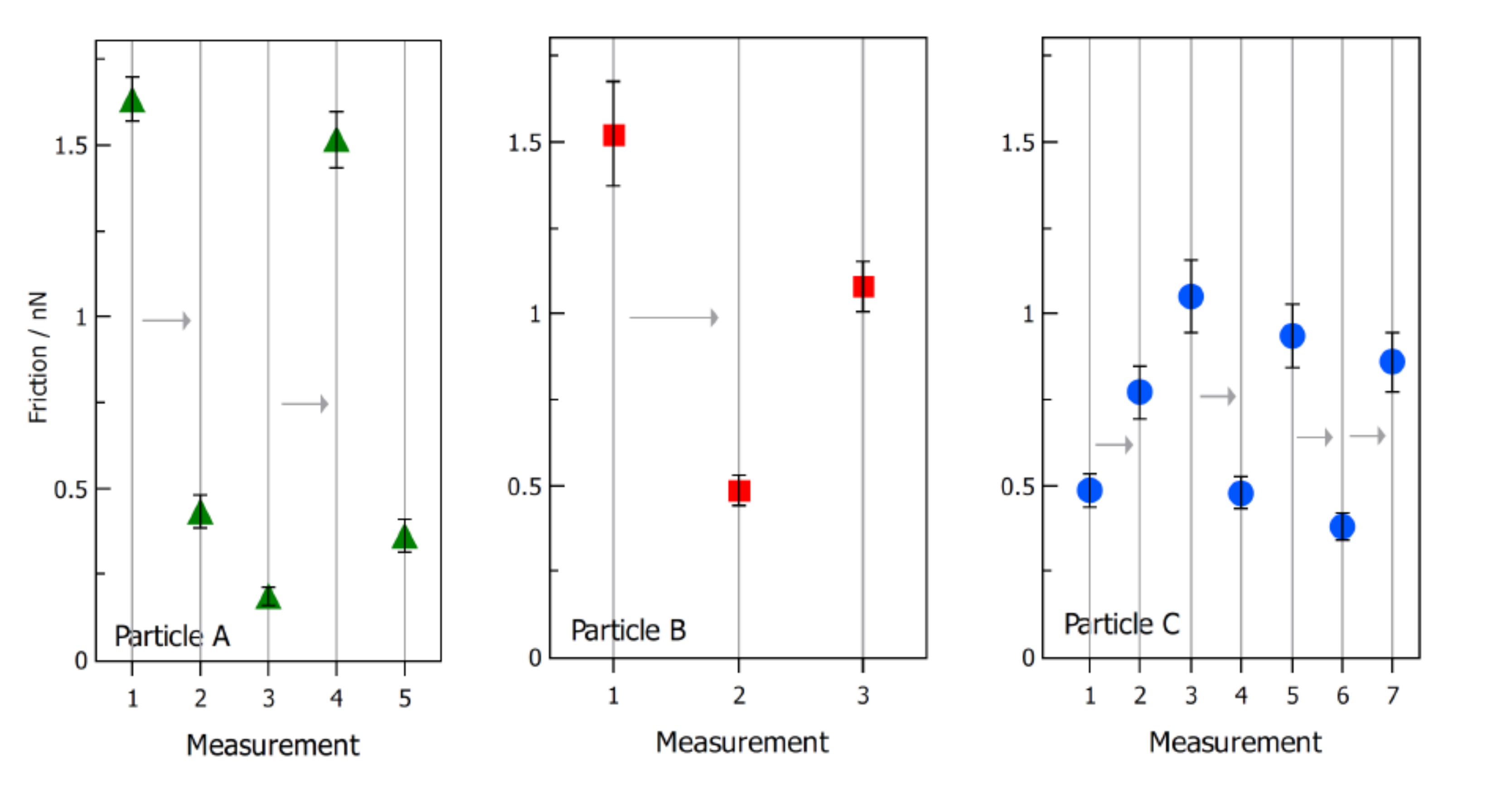}
	\caption{Friction levels observed during sliding of 3 Gold Nanoparticles on HOPG plotted in chronological order. The particle sizes were 3700~nm$^2$ (particle A), 5200~nm$^2$ (particle B), and 12600~nm$^2$ (particle C, cf. Fig\ref{fig:Correlation}). An arrow between data points indicates, that these friction levels were observed within a single friction trace.}
	\label{fig:FrictionTraceSummary}
\end{figure}

Such a behaviour of significant friction fluctuations did not occur for all particles that have been manipulated, but could still be observed for a number of particles, as can be seen from Fig.~\ref{fig:FrictionTraceSummary} where the friction changes found during multiple manipulations of 3 different particles are shown. The different friction levels, which are depicted in chronological order, have either been observed during consecutive manipulations or as distinct friction levels during single manipulation. In each case, the substrate has been checked for surface defects, which in principle could induce similar changes in friction. However, no defects have been observed for the examples in Fig.~\ref{fig:FrictionTraceSummary}. At the same time, these friction changes are reversible, i.e.\ the friction level later sometimes returns to the previous level, within experimental error.
This effect can therefore also not be attributed to irreversible interface changes, such as e.g. accumulation of interface contamination.

Instead, the variations in friction must rather be correlated to intrinsic reversible changes of the interface interaction, which are possible in the context of structural lubricity.
Here, several factors like e.g.\ particle size, particle shape and surface structure are known to be of crucial influence.
However, these parameters are unchanged for each particle during the experiments, and would also not change reversibly.
Among the known factors determining structural lubricity, this leaves only the relative orientation between particle and substrate as a possible candidate to explain the friction fluctuations.

Unfortunately, in our experiments we cannot determine small particle rotation angles very precisely. This is mostly due to the image quality, which is limited by two main factors: First, we need to use a soft contact mode cantilver to achieve sufficient lateral force sensitivity. Such a cantilever is, however, not ideally suited for non-contact operation, which was used to avoid unwanted manipulation. Furthermore, using the same cantilever for multiple manipulations will inevitably increase the tip radius and further decrease the image quality. Consequently, AFM imaging only allows to assert that particle rotation was usually extremely small and even multiple manipulation did not necessarily lead to accumulated rotation. As an example for this behaviour, Fig.~\ref{fig:Correlation}a and b show, that the first manipulation did not induce any substantial rotation, which is instead limited to an angle $ \lesssim 5^{\circ} $. Likewise, the two subsequent manipulations of the same particle only yielded minute rotation angles in the range of the detection threshold (see Fig.~\ref{fig:Correlation}c and d).

Nonetheless, the general occurrence of (albeit very small) particle rotations cannot be ruled out, since our nanomanipulation experiments have been performed by pushing the particles from the side, where the particle is not fully pinned by the AFM tip and freedom of rotation is thus retained. In this case, there are two scenarios how particle rotation can occur: First, it is possible, that the particle is rotated during the contact formation process, when the tip hits the particle. This will lead to changes in friction for two consecutive manipulations of the same particle. Moreover, it is also possible, that the orientation of the particle during manipulation along the tip path is not fully stable so that rotation and thus change in friction can occur suddenly during the movement of the particle. Both effects would indeed be consistent with the experimental results as documented in Fig.~\ref{fig:Correlation} and Fig.~\ref{fig:FrictionTraceSummary}.

In Fig.~\ref{fig:Correlation} e we can first observe a pronounced peak in the lateral force signal, when the AFM-tip reaches the nanoparticle. The occurrence of this friction peak is understandable under the assumption, that the well aligned growth process as shown in Fig.~\ref{fig:SEMAuonHOPG} leads to particle orientation according to a main commensurate direction. At the same time the peak might also be partially attributed to contact ageing processes. However, this friction level is not stable and the lateral force signal immediately drops to appr. 0.5  nN.  The lower friction level after the peak remains constant for appr. 75 nm. After that, a sudden increase in friction can be observed to a level that is kept constant for the remainder of the pushing process and appr. twice as high as the previous level.

Interestingly, a high initial friction peak has also been observed for the second manipulation (Fig.~\ref{fig:Correlation}f) but not for the third one (Fig.~\ref{fig:Correlation}g). This strengthens the assumption, that these peaks are related to the particle orientation, because before all three manipulations, the particles have been at rest for at least several minutes and conventional time dependent contact ageing should thus have affected all three measurements similarly whereas orientation effects might change. While the growth process naturally provides alignment for the first manipulation, the small rotations during manipulations will still allow for the particle to fall back into registry upon tip/particle contact formation but does not neccessarily happen. 

In principle, the idea of rotation induced friction changes can thus already explain a number of tribological effects observed in the measurements.
However, at this point there are still several aspects of the experimental results left that need to be explained.
For example, it is surprising that even multiple particle translation as e.g. shown in Fig. \ref{fig:Correlation} do not induce more substantial rotation.
Additionally, Fig. \ref{fig:FrictionTraceSummary} reveals that friction changes are typically less than one order of magnitude, which is much less than expected by simple application of scaling laws for gold nanoparticles on HOPG \cite{Dietzel2013}.

But even these characteristics can be understood by considering the specific interface conditions of our system. This is demonstrated in the following section, where we have simulated numerically the energy corrugation of gold nanoparticles on HOPG with respect to particle orientation, shape and size. As we will see, friction changes can be understood from a complex and dense pattern of neighbouring commensurate orientations, where substantial rotation is suspended by the energy barriers related to the transition between these orientations.

\section{Modelling Nanoparticle rotation and friction}\label{Section:Modelling}
Interfacial friction of nanocrystals in the superlubric regime involves a complex interplay between the two surfaces and is governed by a variety of parameters like particle size, particle shape, crystalline structure and relative orientations.
Previously, authors have used both numerical and analytical calculations to theoretically describe the interfacial friction.
MD-simulations~\cite{Mueser2001,Filippov2008,deWijn2010, Gnecco2007, Guerra2010, Guerra2016, Varini2014, Pierno2015} can provide a realistic insight into the dynamics and friction of sliding nanoscale objects while analytical calculations as e.g. performed by one of the co-authors\,\cite{deWijn2012} allow for a very precise analysis of specific parameters influencing friction.

\subsection{Stability of incommensurate orientations}

The experimental results in this work can be explained and understood in terms of rotational stability and sliding of nanocrystals~\cite{deWijn2010,deWijn2012}.
It has been shown that only specific incommensurate orienations are stable, and that these orientations can be found from the interfacial energy, as they coincide with maxima in the surface corrugation as a function of orientation.
The number of these stable incommensurate orientations is large, growing with the square root of the number of interfacial atoms.
Because the lattice parameters of gold and graphite do not match, there are no fully commensurate configurations, but there are several trivial and nontrivial pseudo-commensurate orientations.

In order to study this in more detail for the case of gold on graphite, we have calculated numerically corrugations as a function of orientation for various shapes and sizes, and we interpret this in the context of previous theoretical results for stability of the orientations~\cite{deWijn2010}.
We use a two-dimensional hexagonal substrate with sinusoidal potential energy and a maximum corrugation of $E_0$, i.e. $V(x,y) = (2/9) E_0 \sum_{i=1}^3 \cos(4 \pi (x,y) \cdot \vec{e}_i /(3 a))$, with $\vec{e}_1=(1,0), \vec{e}_2= (\frac12,\frac12\sqrt{3}), \vec{e}_3=(-\frac12,\frac12\sqrt{3})$, and $a$ the inter-atomic distance of graphite, 0.142~nm.
On this substrate we place a rigid body of atoms arranged in a triangular lattice with the inter-atomic distance of the (111) surface of gold, 0.276~nm.  We rotate the crystals and sum up the potentials of all atoms numerically.
In order to determine the corrugation, we move the crystal around on the surface to find the highest and lowest potential energies.

\begin{figure}
\includegraphics[angle=270,width=\linewidth,keepaspectratio]{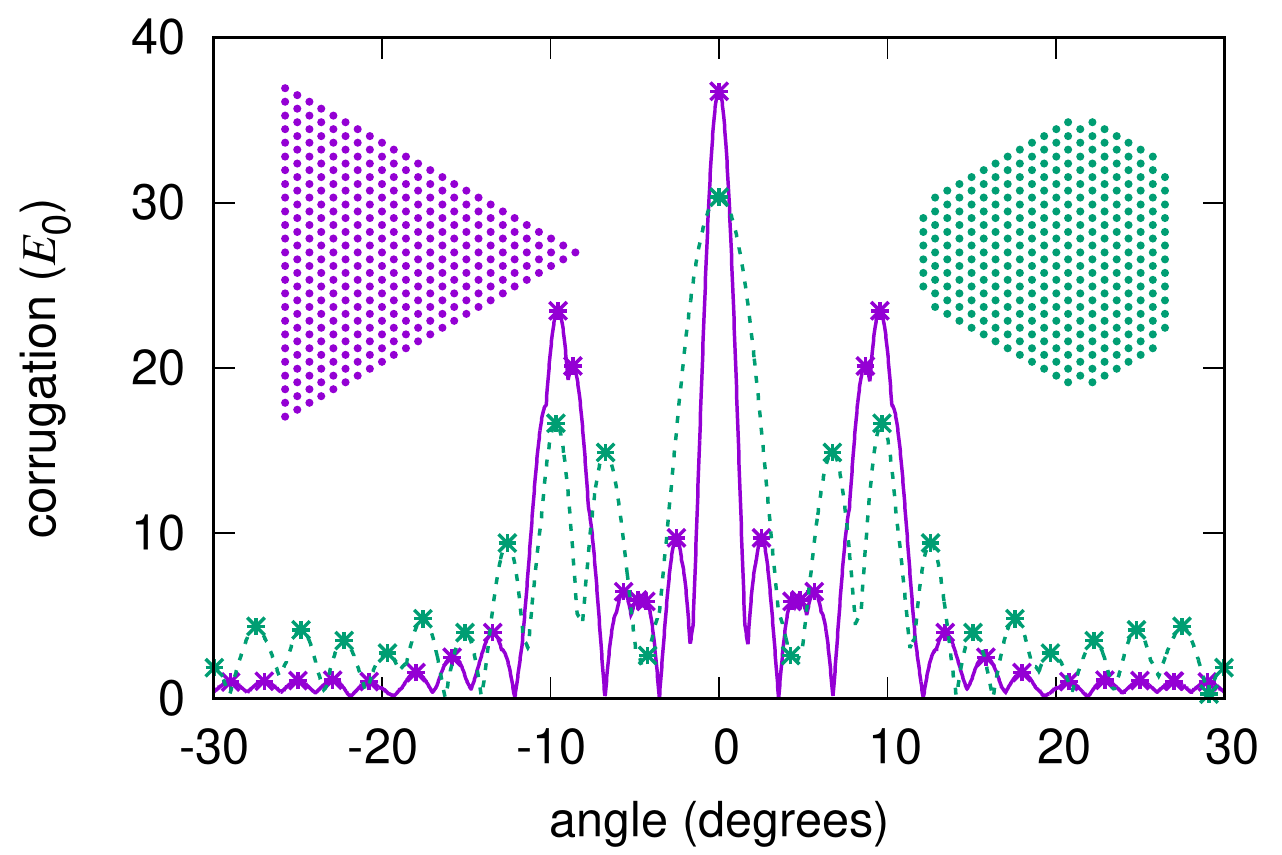}
\caption{An example of the corrugation as a function of orientation angle for a triangular crystal of 325 atoms and a crystal with rounded corners of 316 atoms. The maxima are indicated with stars. Each of these corresponds to an orientation that is stable for some positions on the substrate~\cite{deWijn2010}.  The friction for these orientations is approximately linearly dependent on the corrugation.\label{fig:simple}}
\end{figure}

\begin{figure}
\includegraphics[width=\linewidth,keepaspectratio]{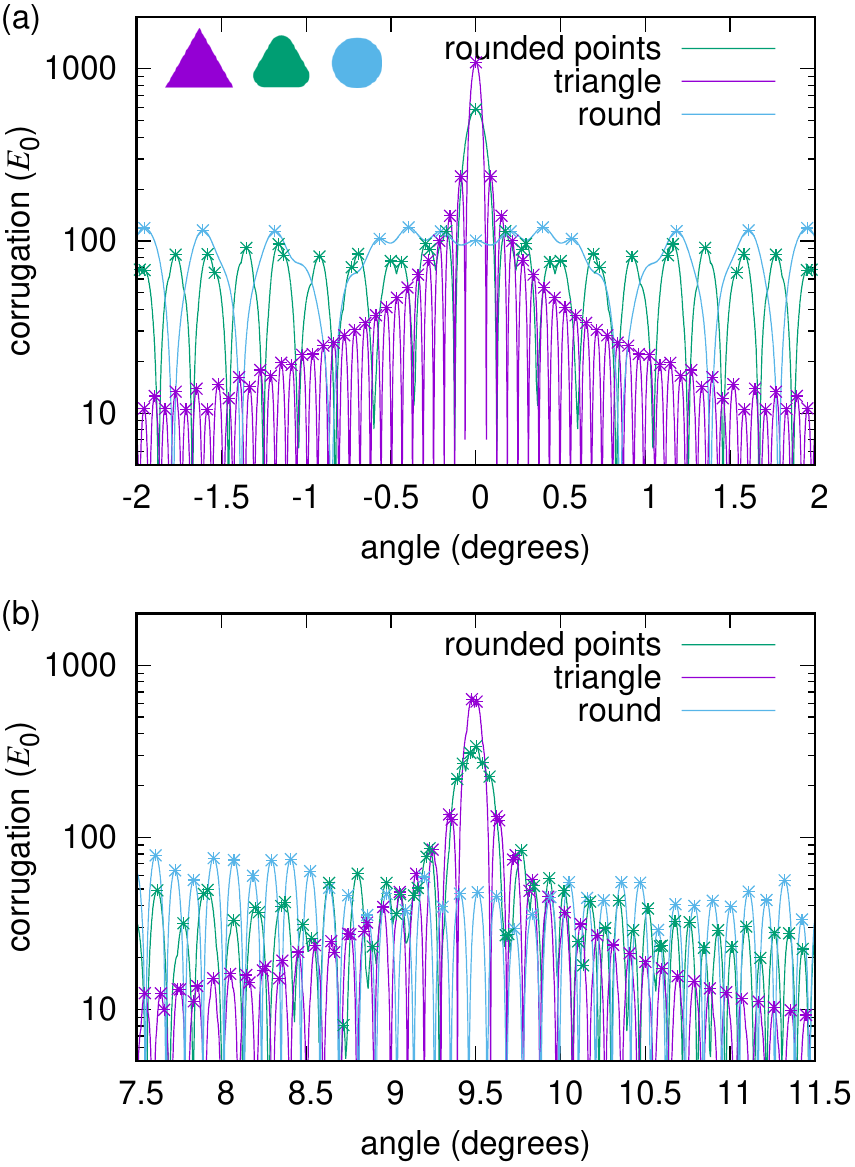}
\caption{The corrugation as a function of the orientation for three different shapes close to the commensurate orientations for very large crystals of contact area between 253500 and 253900 atoms.
Subfigure (a) shows the obvious commensurate orientation, while subfigure (b) shows the non-trivial commensurate orientation.
\label{fig:largecrystal}}
\end{figure}

We illustrate an example of the maxima and corresponding possible stable orientations for two relatively small crystals in figure~\ref{fig:simple}.
As the particle size increases, the number of stable orientations also increases~\cite{deWijn2012}, linearly with the diameter.
This is illustrated in Fig.~\ref{fig:largecrystal}, where the corrugation is shown for several very large crystals of about $2.5\cdot10^5$ atoms in a small range.
There are two orientations with very large corrugation, corresponding to the trivial and non-trivial pseudo-commensurate orientations found in the analytical calculations~\cite{deWijn2012}.

\begin{figure}
\includegraphics[angle=270,width=\linewidth,keepaspectratio]{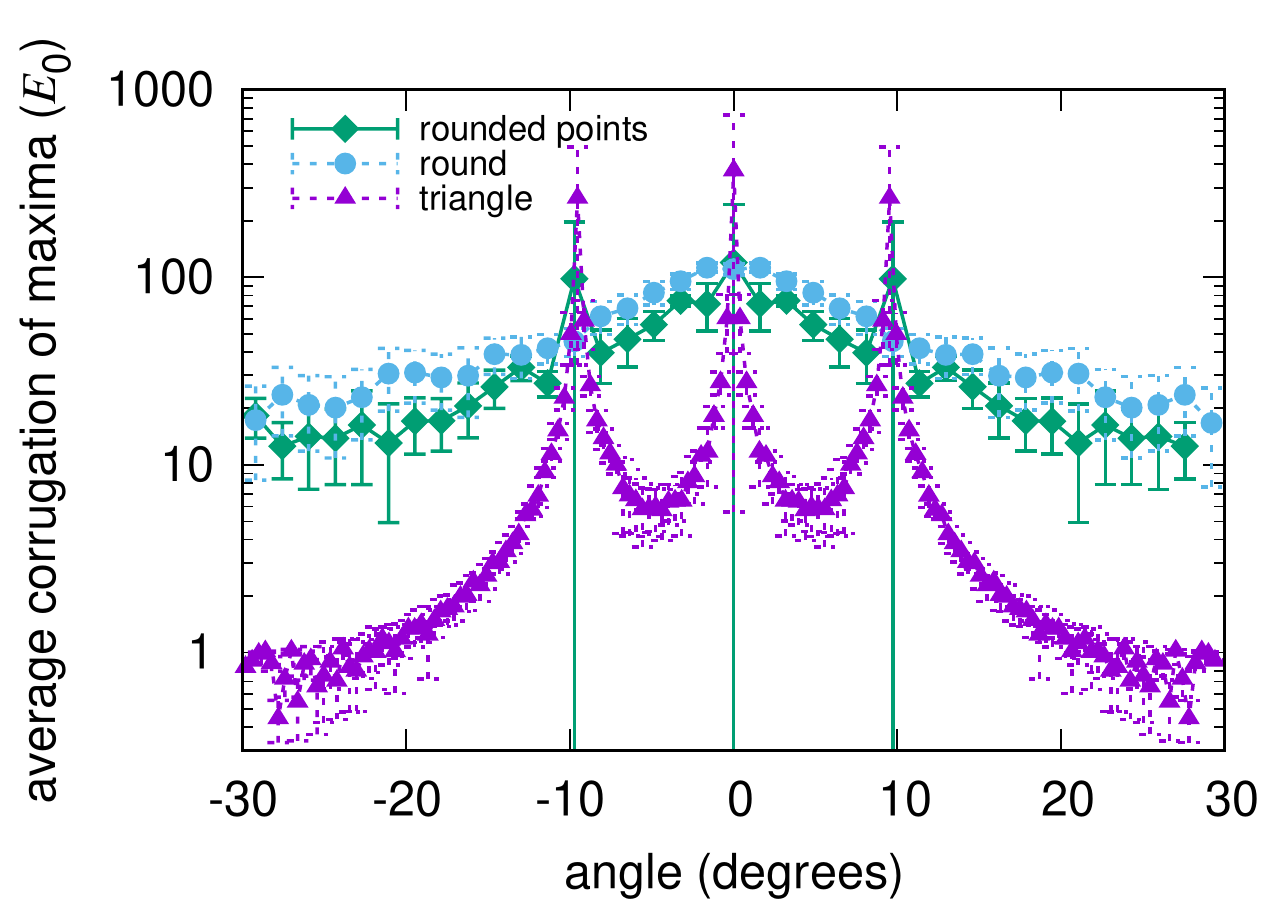}
\caption{The average height of the maxima in the corrugation as a function of orientation angle for three crystals of different shape but similar contact area of between 253500 and 253900 atoms. The maxima have been averaged in bins with a width of about 0.81$^\circ$.  The errorbars indicate the standard deviation in the maxima. For these nanocrystals, there are a total of between 500 and 1000 maxima in the corrugation as a function of the orientation.\label{fig:statistics}}
\end{figure}

In general, even nearby maxima can vary in height substantially.
This can be seen from Fig.~\ref{fig:largecrystal}, where there are many side maxima of different height.
This is important, because the corrugation is also directly related to the friction.  At low sliding speeds (such as those in the experiments), there is generally a linear dependence.
We therefore investigate the typical height of the maxima, and the standard deviation in the heights of the maxima.
This is shown in Fig.~\ref{fig:statistics} for some very large crystals and three different shapes.
One can see from this figure that the standard deviation is significant, and thus nearby maxima can differ by a factor of 2 or more, especially close to the pseudo-commensurate orientations.
In fact, near the pseudo-commensurate orientations, the standard deviation is extremely high, despite the small bin size we have employed ($1.6^\circ$ for the round and rounded crystals, and $0.8^\circ$ for the perfect triangular crystal).

For sufficiently large crystals the different stable orientations are so close together that they could not be distinguished in a topography scan.  But, because of the changes in corrugation, the friction during sliding still changes significantly.
We therefore conclude that small rotations, between nearby stable orientations, can account for the variation in the friction that is observed in the experiments.
Moreover, rotations are reversible, unlike other candidate mechanisms that might change the interaction between the nanocrystal and substrate.
Given that sometimes the friction changes also reverse completely in the experiments, small rotations are therefore the most likely explanation for the behaviour that is observed.

\subsection{Relation between shape and scaling}

\begin{figure}
\includegraphics[angle=270,width=\linewidth,keepaspectratio]{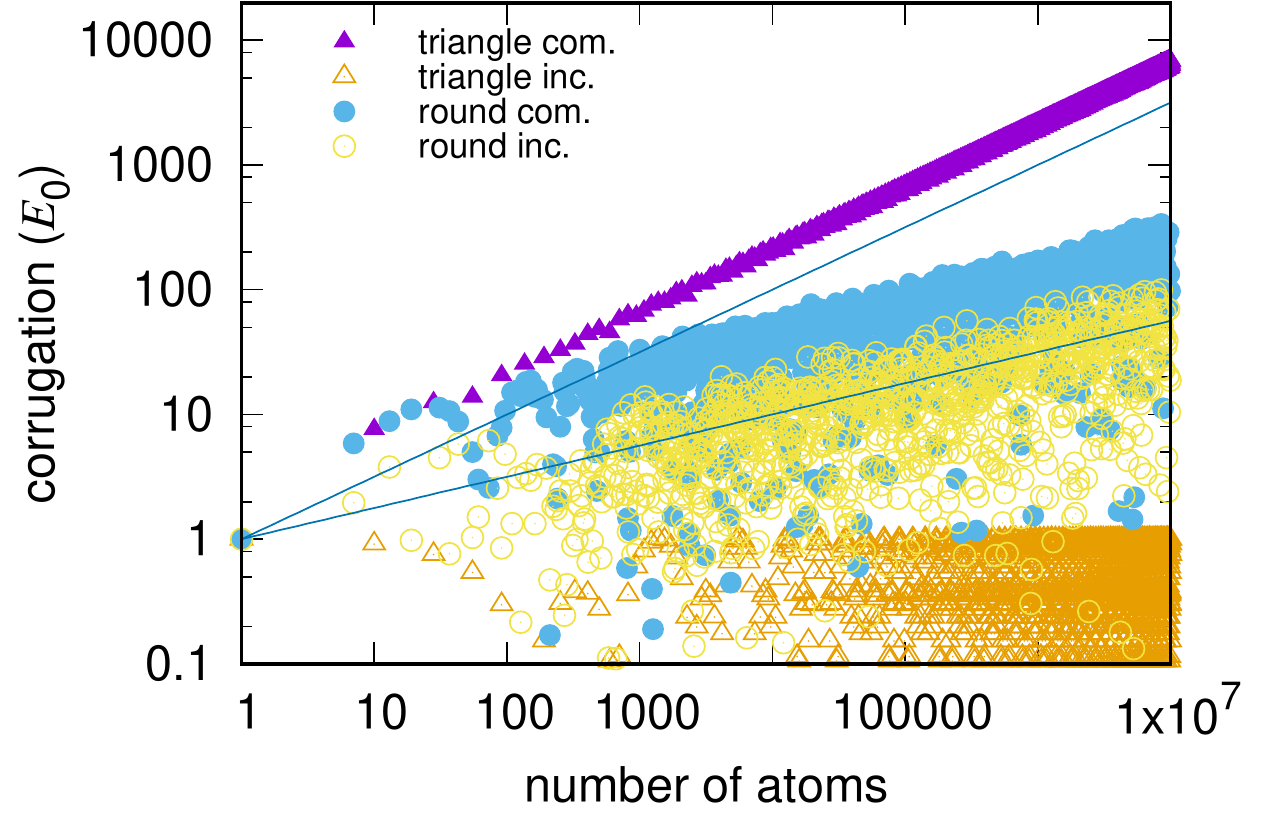}
\caption{The corrugation as a function of the crystal size for triangular and round shapes and both pseudo-commensurate ($0^\circ$) and incommensurate ($30^\circ$) orientation.  The lines indicate scaling with exponent $\frac12$ and $\frac14$.}\label{fig:scaling1}
\end{figure}

\begin{figure}
\includegraphics[angle=270,width=\linewidth,keepaspectratio]{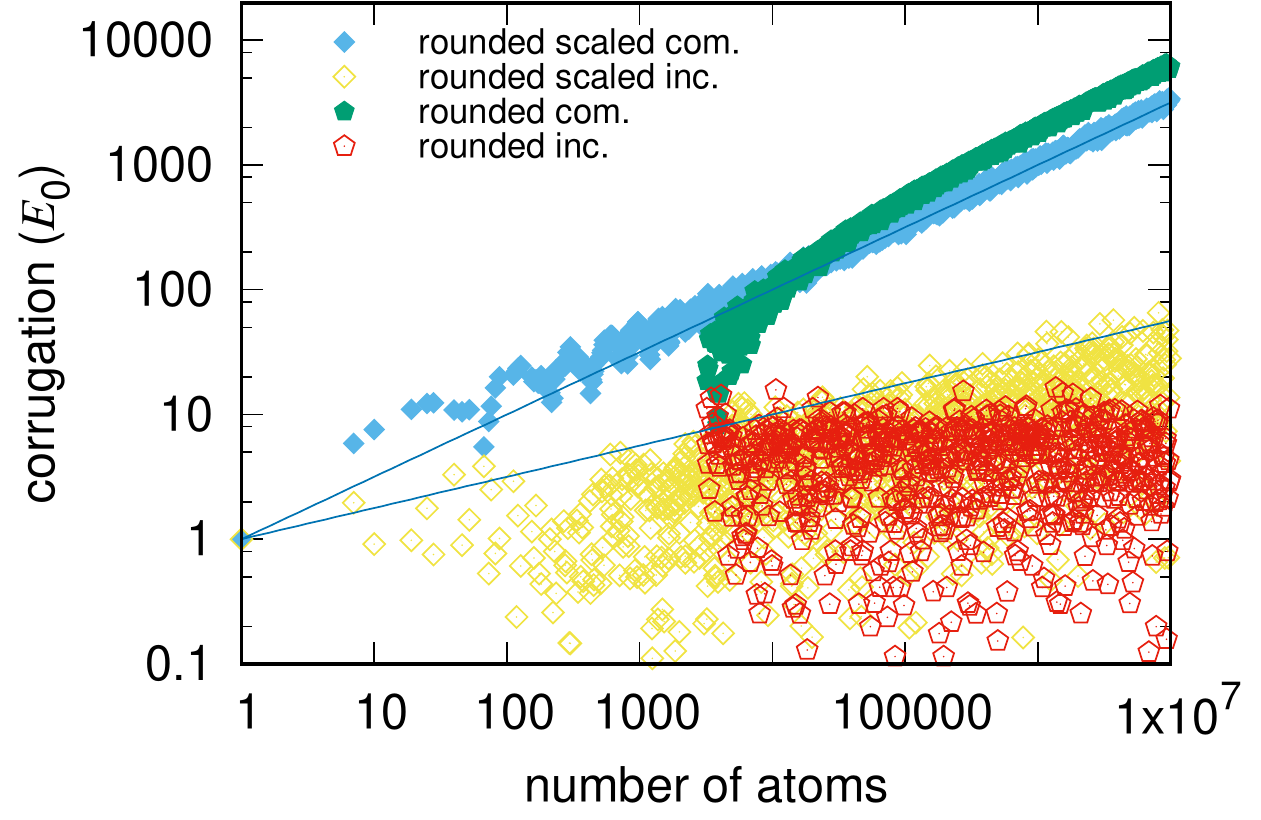}
\caption{The corrugation as a function of the crystal size for crystals with rounded corners and both pseudo-commensurate ($0^\circ$) and incommensurate ($30^\circ$) orientation.  The lines indicate scaling with exponent $\frac12$ and $\frac14$.  The scaling depends on how the rounded corners are treated.  If the crystal shape is scaled up as a whole, then the scaling is dominated by the rounded corners and similar to that of a round crystal. If the radius of curvature of the corners is kept constant instead, the scaling follows that of a perfectly triangular crystal.}
\label{fig:scaling2}
\end{figure}

From Fig.~\ref{fig:statistics} one can also see that the shape of the crystal has a significant impact.
The extremes of the pseudo-commensurate and incommensurate orientations are washed out by the more complex shapes.
There are significant quantitative differences and also differences in the dependence of the friction on the contact size~\cite{deWijn2012}.
We also study numerically the scaling of the corrugation (and friction) with the size of the contact for a number of different shapes of crystals.
Results of this are shown in Figures~\ref{fig:scaling1} and~\ref{fig:scaling2}.

For the perfect crystals in pseudo-commensurate and incommensurate orientations, the corrugation scales with the contact area with exponents $\frac12$ and $0$ respectively.
For the rounded crystals, this distinction between pseudo-commensurate and incommensurate is reduced, and both scale with exponent $\frac14$, though there is still a quantitative difference of an order of magnitude.
To better understand the effect of shape and scaling, we compare triangular crystals with rounded corners that either do or do not scale up with particle size.  In the first case, the shape of the crystal is kept constant, while in the second case the radius of curvature of the corners is constant.
This comparison is shown in Fig.~\ref{fig:scaling2}.
For large enough contact area, the crystals with corners of constant radius of curvature show the same scaling as triangular crystals (with exponents 0 and $\frac12$), though the prefactor for the incommensurate orientations differs.
The crystals with constant shape, however, scale differently from both round and perfectly triangular crystals.  The exponent for incommensurate crystals is that of round crystals (i.e.\ $\frac14$), but for the pseudo-commensurate orientation, the scaling is that of a triangular crystal ($\frac12$).
It is thus the straight edge, when present, dominates the scaling for pseudo-commensurate orientations with exponent $\frac12$, while the round edges dominate for the incommensurate orientations, giving exponent $\frac14$.
We conclude from this that in experiments, it is not always easy to distinguish between pseudo-commensurate and incommensurate orientations, as the crystals do not usually have a perfect shape.  

\section{Conclusion and Outlook}

In this paper we have analyzed friction fluctuations observed during lateral manipulation of gold nanoparticles on HOPG. These fluctuations have been associated with particle rotation, where changing levels of commensurability between substrate and slider can induce variations in the energy landscape experienced by the moving nanoparticles. 

The principle importance of (pseudo-)commensurate orientations for the Au/HOPG systems has been previously predicted by analytical calculations~\cite{deWijn2012} and is corroborated in this work by the uniform alignment, that was found for gold nanoparticles on HOPG grown by thermal evaporation (Fig. \ref{fig:SEMAuonHOPG}). 

To further explain these friction variations, we have performed detailed theoretical calculations, which have revealed an intricate succession of multiple minima and maxima of the energy as a function of the particle orientation.  Each of these minima and maxima corresponds to a stable orientation with a different friction.
For nearby orientations we have computed the variation in the energy barriers against sliding, which are directly related to the friction.
We find that small rotations between these states can lead to changes in friction substantial enough to explain the fluctuations in the friction found in the experiments.
Here, especially the shape of the particles is of crucial importance.
By comparing the energy landscapes of triangular, rounded, and fully round particles we have shown that the variation in the friction is only compatible with the experimental results if the crystals are not perfectly triangular.
In this case, especially the height variations in the energy landscape for different incommensurate orientations match the fluctuations found in the experimental results.   

Particle rotation during nanomanipulation experiments was typically limited to angles of only a few degrees, even after pushing the nanoparticles multiple times with several changes of friction between higher and lower levels. This stability of particle orientation is in accordance with the existence of an effective energy barrier preventing more substantial rotation. On the other hand, the close proximity of different incommensurate orientations still allows for frequent friction changes, even with nanoparticle rotation limited to extremely small angles.

Based on these results, further experiments should now be performed to analyse the complex interdependence between shape, orientation and friction in more detail. In this context especially inducing larger rotations will be interesting. Although the close proximity of the different incommensurate orientations makes it impossible to distinguish these, it will still be highly interesting to see to what extend the principle shape of the curves from Fig. \ref{fig:statistics} can be reproduced.

\section*{Acknowledgements}
Financial support was provided by the German Research Foundation (Project DI917/5-1) and in part by COST Action MP1303. ASdW acknowledges support from the Swedish Research Council (Vetenskapsr\aa{}det).

\section*{References}

\bibliographystyle{unsrt}

\end{document}